\documentclass[12pt]{amsart}

\newtheorem{theorem}{Theorem}[section]
\newtheorem{lemma}[theorem]{Lemma}

\theoremstyle{definition}
\newtheorem{definition}[theorem]{Definition}

\theoremstyle{remark}
\newtheorem{remark}[theorem]{Remark}
\numberwithin{equation}{section}

\def\RE{\mathbb R}
\def\CO{{\mathbb C}}

\def\p{\par\noindent}

\def\sgn{\text{\rm sgn}}

\begin{document}

\title[Wave equation with concentrated nonlinearities] {Wave Equations
  with concentrated nonlinearities} 
 \author{Diego Noja}
\address{Dipartimento di Matematica e Applicazioni, Universit\`a Di
  Milano-Bicocca, I-20126, Milano, Italy}
\email{noja@matapp.unimib.it} 
\author{Andrea Posilicano}
\address{Dipartimento di Scienze Fisiche e Matematiche, 
Universit\`a dell'Insubria,
  I-22100 Como, Italy} \email{posilicano@uninsubria.it}

\begin{abstract} 
In this paper we address the problem of wave dynamics in presence of
concentrated nonlinearities. Given a vector field $V$ on an open
subset of $\CO^n$ and a discrete set $Y\subset\RE^3$ with $n$ elements, we
define a nonlinear 
operator $\Delta_{V,Y}$ on $L^2(\RE^3)$
which coincides with the free Laplacian when restricted to regular
functions vanishing at $Y$, and which
reduces to the usual Laplacian with point interactions placed at $Y$
when $V$ is linear and is represented by an Hermitean matrix.
We then consider the nonlinear wave equation $\ddot
\phi=\Delta_{V,Y}\phi$ and
study the corresponding Cauchy problem, giving an existence and
uniqueness result in the case $V$ is Lipschitz. The solution of such a problem is explicitly expressed in
terms of the solutions of two Cauchy problem: one relative to a free
wave equation and the other relative to an inhomogeneous
ordinary differential equation with delay and principal part 
$\dot\zeta+V(\zeta)$.
Main properties of the solution are 
given and, when $Y$ is a singleton, 
the mechanism and 
details of blow-up are studied.
\end{abstract}

\maketitle

\section{Introduction}
In recent times a great work has been devoted to the analysis of
nonlinear wave equations. 
Among the more interesting themes, there are global
existence, presence of blow-up solution and characterization of their
lifespan (see e.g. \cite {[alinhac]}, \cite{[merle1]}, \cite{[caff]},
\cite{[Horm]}, \cite{[john]}, \cite{[merle2]}, \cite{[struwe]} and
references 
therein).  These issues are
usually quite difficult to analyze, due to the the scarcity of
information about exact solutions of nonlinear wave equations. In
this paper we study a class of wave equations about which information
on exact solutions is relatively easy to obtain.  This class is 
characterized by a so called concentrated
nonlinearity, modellized as a nonlinear point interaction in some
fixed finite set of points.  To be more precise, we will study abstract wave
equations of the form $\ddot\phi=\Delta_{V,Y}\phi$, where
$\Delta_{V,Y}$ is a nonlinear operator on
$L^2(\RE^3)$ which coincides
with the free Laplacian when restricted to regular functions vanishing
at the point of $Y=\left\{y_1,\dots,y_n\right\}$, a discrete subset of
$\RE^3$. $V\equiv(V_1,\dots,V_n)$ is a vector
field on $\CO^n$ which 
is related to the beaviour at $Y$ of the functions $\phi$ belonging to
the domain of $\Delta_{V,Y}$ by
$$
\phi(x)=\frac{\zeta_j^\phi}{4\pi|x-y_j|}+V_j(\zeta^\phi)+O(|x-y_j|)\,,\quad
1\le j\le n\,,\quad x\to y_j\,.
$$ 
The action of $\Delta_{V,Y}$ can be then defined in a suggestive way by
$$
\Delta_{V,Y}\phi:=\Delta\phi+\sum_{1\le j\le n}\zeta^\phi_j\delta_{y_j}\,,
$$
where $\delta_y$ is the Dirac mass at $y$. We refer to Section 2 for the
precise definitions. Such a nonlinear operator reduces to the
self-adjoint operator given by the Laplacian with $n$ point
interactions (see \cite{[AGHH]}, \cite{[AK]}) in
the case $V(\zeta)=\Theta\zeta$, $\Theta$ an Hermitean matrix.\par
In Section 3 we then turn to the problem of existence and uniqueness
of the Cauchy problem for the nonlinear wave equation
$\ddot\phi=\Delta_{V,Y}\phi$. The analogous problem for the nonlinear Schr\"odinger
equation $i\dot\psi=-\Delta_{V,Y}\psi$ was studied in \cite{[ADFT1]} and
\cite{[ADFT2]} in the case of particular nonlinearities of the kind 
$V_j(\zeta)=\gamma_j|\zeta_j|^{2\sigma_j}\zeta_j$, $\gamma_j\in\RE$,
$\sigma_j\ge 0$, whereas the wave equation case was studied, when $V$
is linear, in \cite{[NP1]}, \cite{[NP2]}, \cite{[NP3]}, \cite{[BNP]} in the
case $Y$ is a singleton and in \cite{[KP]} in the general case. The 
nonlinear wave equation case was
instead totally unexplored. Thus is Theorem 3.1 we provide an existence
and uniqueness result in the case $V$ is Lipschitz. The strategy of
the proof follows the lines of the linear case with the complication
due to the lack of a general existence theorem in this singular
situation. Similarly to the linear case, the main result is
the relation between the equation $\ddot\phi=\Delta_{V,Y}\phi$ 
and a coupled system constituted of an ordinary wave equation with
delta-like sources and an inhomogeneous ordinary differential 
equation with
delay driven by the vector field $V$. 
This delayed equation controls the dynamics of the coefficients 
$\zeta^\phi$, and an almost complete decoupling is
achieved in that it is possible to get the solution of the
problem when the retarded Cauchy problem for the $\zeta^\phi$'s (depending in
a parametric way from the initial data fo the field $\phi$) is
solved, apart a term which is the free wave evolution of the initial
data. A similar situation appears (for the particular nonlinearities
indicated above) in the Schr\"odinger case
(see \cite{[ADFT1]}) where however, due to infinite speed of
propagation of the free Schr\"odinger equation,
an integral Volterra type equation substitutes the ordinary
differential equation.
\par In the case the vector field $V$ is of gradient type, a
conserved energy for the dynamics exists (see Lemma 3.7) and this
provides criteria for global existence (see Theorem 3.8). 
When there is no global
existence, the problem of the characterization of blow-up solutions and
their blow-up rates arises. In the special case
where the singularity is at
only one point $y$, a detailed study is possible (see Section 4). 
The key remark is that the
inhomogeneous term in the equation for the $\zeta^\phi$'s 
is bounded and continuous and there exist simple autonomous first order
differential equations the solution of which provide supersolutions
and subsolutions by means
of differential inequalities. This allows to prove in many cases
existence in the large or on the contrary blow-up of the solutions
together, in the latter case,
with an extimate of their lifespan. A typical example is a
power law nonlinearity, where explicit calculations are given and in
particular for the quadratic nonlinearity, which leads to an equation of
Riccati type.

\section{Nonlinear point interactions}
Given the vector field $V:A_V\subseteq\CO^n\to\CO^n$, $A_V$ open, 
and a discrete set $Y\subset\RE^3$, $Y=\left\{y_1,\dots,y_n\right\}$, 
we give the definition of a nonlinear 
operator $\Delta_{V,Y}$ on $L^2(\RE^3)$
which reduces to the usual Laplacian with point interactions at 
$Y$ when $V$ is linear and is represented by an Hermitean matrix.
 
\begin{definition}  We define the nonlinear subset $D_{V,Y}$ of 
$L^2(\RE^3)$ by the set of $\phi\in L^2(\RE^3)$ for which there exists
an $n$-uple of complex numbers
$\zeta^\phi=(\zeta^\phi_1,\dots,\zeta^\phi_n)\in A_V$ 
such that $$\phi_{reg}\in\bar H^2(\RE^3):=\left\{f\in
L^2_{loc}(\RE^3)\,:\, \nabla f \in L^2(\RE^3)\,,\quad 
\Delta f\in L^2(\RE^3)\right\}\,,$$ 
where
$$ 
\phi_{reg}:=\phi-\sum_{1\le j\le n} \zeta^\phi_j G_j\,,
\qquad
G_j(x):=\frac{1}{4\pi|x-y_j|}\,,
$$
and moreover the following nonlinear boundary conditions holds true
at $Y$
\begin{equation}\label{bc}
\lim_{x\rightarrow y_j}\left(\phi (x) -\zeta^\phi_j G_j(x)\right)=
V_j(\zeta^\phi)\,, \qquad 1\le j\le n\,,
\end{equation}
where $V(\zeta)\equiv(V_1(\zeta),\dots,V_n(\zeta))$.
The action of
$$
\Delta_{V, Y} : D_{V,Y}\subset L^2(\RE^3) \rightarrow L^2(\RE^3)
$$
is then given by
$$
\Delta_{V, Y}{\phi} := \Delta{\phi}_{reg}\,. 
$$
\end{definition}
The set $Y$ is the singular set of the point interaction. It is the
set where the elements of the domain of $\Delta_{V, Y}$  do not 
belong to $\bar H^2(\RE^3)$, or better, since
$\bar H^2(\RE^3)\subset C_b(\RE^3)$, where they are unbounded.  

Let us define, for any $z\in \CO\backslash(-\infty,0]$,
$$
G_i^z(x):=\frac{e^{{-\sqrt z}\,|x-y_i|}}{4\pi|x-y_i|}\,,\quad
\text{\rm Re$\sqrt z>0$}\,,
$$
and
$$ 
(M_Y(z))_{ij}:=(1-\delta_{ij})\, G_i^z(y_j) 
\,,\quad\langle G_Y^z,\phi\rangle_i:=\langle G_i^z,\phi\rangle\,.
$$
Then one has the following
\begin{lemma}For any $z\in\CO\backslash(-\infty,0]$ such that the function 
$$
\Gamma_{V,Y}(z):A_V\to\CO^n\,,\quad
\Gamma_{V,Y}(z):=V+\frac{\sqrt z}{4\pi}-M_Y(z)
$$
has an inverse, 
the nonlinear resolvent of $\Delta_{V,Y}$ is given by
$$
(-\Delta_{V,Y}+z)^{-1}\phi=(-\Delta+z)^{-1}\phi+\sum_{1\le i\le n}
(\Gamma_{V,Y}(z)^{-1}\langle G_Y^{\bar z},\phi\rangle)_i 
G_i^z\,.
$$
\end{lemma}
\begin{proof} We need to solve the equation
$(-\Delta_{V,Y}+z)\psi=\phi$. By the definition of $\Delta_{V,Y}$ 
one has
$$
\psi_{reg}=(-\Delta+z)^{-1}\phi-z\sum_{1\le i\le n}\zeta^{\psi}_i
(-\Delta+z)^{-1}G_i
$$
and 
$$
\psi_{reg}(y_j)=\left((V-M_Y(0))(\zeta^\psi)\right)_j=
\langle G_j^{\bar z},\phi\rangle-z\sum_{1\le i\le n}\zeta^{\psi}_i
\langle G_j^{\bar z},G_i\rangle\,.
$$
Since $z\langle G_j^{\bar z},G_i\rangle=(M_Y(0)-M_Y(z))_{ij}$ and 
$z\langle G_i^{\bar z},G_i\rangle=1/4\pi\sqrt z$,
one obtains
$$\zeta^\phi=\Gamma_{V,Y}(z)^{-1}\langle G_Y^z,\phi\rangle$$
so that 
\begin{align*}
\psi=&(-\Delta+z)^{-1}\phi+
\sum_{1\le i\le n}(\Gamma_{V,Y}(z)^{-1}\langle G_Y^{\bar z},\phi\rangle)_i
(G_i-z(-\Delta+z)^{-1}G_i)\\
=&(-\Delta+z)^{-1}\phi+
\sum_{1\le i\le n}(\Gamma_{V,Y}(z)^{-1}\langle G_Y^{\bar z},\phi\rangle)_i
G_i^z\,.
\end{align*}
\end{proof}
\begin{remark} The nonlinear resolvent
$R_{V,Y}(z):=(-\Delta_{V,Y}+z)^{-1}$ satisfies the nonlinear resolvent
identity 
$$
R_{V,Y}(z)=R_{V,Y}(w)(1-(z-w)R_{V,Y}(w))\,.
$$
Thus $\Delta_{V,Y}$ can be alterantively defined as 
$$\Delta_{V,Y}\phi:=(-R_{V,Y}(z)^{-1}+z)\phi
=\Delta\phi_z+z\sum_{1\le j\le n}\zeta_j^\phi G_j^z$$ on 
\begin{align*}
&D_{V,Y}:=\text{\rm Range}(R_{V,Y}(z))\\
&=\left\{\phi\in L^2(\RE^3)\,:\, 
\phi=\phi_z+\sum_{1\le j\le n}\zeta_j^\phi G_j^z,\quad \phi_z\in
H^2({\RE^3}),\right.\\
&\qquad\left.\Gamma_{V,Y}(z)\zeta^\phi=(\phi_z(y_1),\dots,\phi_z(y_n))
\right\}\,,
\end{align*} 
the definition being $z$-independent. In the above definition 
$H^2({\RE^3}):=\bar
H^2({\RE^3})\cap L^2(\RE^3)$ denotes the usual Sobolev space of index
two. For future convenience we also introduce the Sobolev spaces of index
one: 
$$\bar H^1(\RE^3)
:=\{f\in L_{loc}^2(\RE^3)\,:\, \nabla f\in
L^2(\RE^3)\}\,,$$
and $H^1(\RE^3):=\bar
H^1({\RE^3})\cap L^2(\RE^3)$.\par
Lemma 2.2 also shows that when $V$ is linear and is represented by an 
Hermitean matrix $\Theta$, the linear operator $\Delta_{\Theta, Y}$
coincides with the self-adjoint operator giving the usual Laplacian with 
$n$ point interactions placed at $Y$ (see \cite{[AGHH]}, \cite{[AK]}). 
\end{remark}
The form domain of the operator 
$\Delta_{\Theta, Y}$, which we denote by $\dot D_Y$, is the 
set of $\phi\in L^2(\RE^3)$ for which there exists
an $n$-uple of complex numbers $\zeta^\phi=(\zeta^\phi_1,\dots,\zeta^\phi_n)$ 
such that $\phi_{reg} \in \bar H^1(\RE^3)
$, where $\phi_{reg}\in 
L_{loc}^2(\RE^3)$ is defined as before. Note that here 
no restriction at all is imposed on the vector 
$\zeta^\phi$ so that $D_{V,Y}\subset \dot D_Y$. The quadratic form corresponding to the linear 
operator $-\Delta_{\Theta, Y}$ is then given by 
$$
{\mathcal F}_{\Theta,Y}(\phi)=\|\nabla\phi_{reg}\|_{L^2}^2  -(M_Y\zeta^\phi,\zeta^\phi)+
(\Theta\zeta^\phi,\zeta^\phi)
$$ 
(see \cite{[T]}), where $(\cdot,\cdot)$ denotes the usual Hermitean scalar product on
$\CO^n$ and $M_Y$ is the symmetric matrix $M_Y:=M_Y(0)$ .

\section{Existence and Uniqueness}

\begin{theorem} Let $V:A_V\subseteq\CO^n\to\CO^n$ be Lipschitz, let 
$\phi_0 \in D_{V,Y}$ and $\dot\phi_0 \in
\dot D_Y$. Let $\zeta(t)$, $t\in(-T,T)$, 
be the unique maximal solution of the Cauchy problem with delay 
\begin{align}\label{delay}
&\frac{\sgn(t)}{4\pi}\,\dot\zeta_j(t)+V_j(\zeta(t)) =\nonumber \\
&\sum_{i\not=j}\frac{\theta(|t|- |y_i-y_j|)}{4\pi|y_i-y_j|}\, 
\zeta_i(t-\sgn(t)\,|y_i-y_j|)
+ \phi_f(t,y_j)\,,\\
&\zeta(0)=\zeta^{\phi_0}\,,\qquad 1\le j\le n\,,\nonumber
\end{align}
where $\theta $ denotes the Heaviside function and  
$\phi_f$ is the unique solution of the Cauchy problem
\begin{align}\label{free}
&\ddot{\phi}_f(t)  =   \Delta \phi_f(t)\nonumber \\
&\phi_f(0) =  \phi_0 \\
&\dot \phi_f(0) = \dot\phi_0 \nonumber
\end{align}
Defining, given $s\in\RE$,   
\begin{align*}
&\phi(t,x):= \phi_f(t-s,x)\\ &+ 
\sum_{1\le j\le n}\frac{\theta(|t-s|-|x-y_j|)}{4\pi|x-y_j|}\, 
\zeta_j((t-s)-\sgn(t-s)\,|x-y_j|)
\,,
\end{align*}
one has 
$$
\phi(t)\in D_{V,Y}\,,\quad \dot\phi(t)\in \dot D_Y\,,\quad
\zeta^{\phi(t)}=\zeta(t-s)\,,\quad\zeta^{\dot\phi(t)}=\dot\zeta(t-s)$$
for all
$t\in(-T+s,T+s)$ and 
$\phi$ is the unique strong solution of the Cauchy problem  
\begin{align}\label{CP}
&\ddot{\phi}(t) =   \Delta_{V,Y}\, \phi(t)\nonumber \\
&\phi(s) = \phi_0 \\
&\dot\phi(s)  =  \dot\phi_0\,.\nonumber
\end{align}
Moreover, defining the nonlinear map 
$$
U_{V,Y}(t):D_{V,Y}\times \dot D_Y\to D_{V,Y}\times \dot D_Y\,,\qquad
t\in(-T,T)\,,
$$
$$U_{V,Y}(t)(\phi_0,\dot\phi_0):=(\phi(t+s),\dot\phi(t+s))\,,$$
one has, for any
$t_1,t_2\in (-T,T)$ with $t_1+t_2\in(-T,T)$, the group property  
\begin{equation}\label{flow}
U_{V,Y}(t_1)U_{V,Y}(t_2)=U_{V,Y}(t_1+t_2)\,.
\end{equation}
\end{theorem}
We premise to the proof some preparatory lemmata:
\begin{lemma}
Let $\xi:(a,b)\to \CO$. Then 
$$\xi\in L^2_{loc}(a,b)\quad\iff\quad
\forall\,t\in(a,b)\,,\quad\psi(t)\in H^2(\RE^3)\,,
$$
$$
\xi\in L^2_{loc}(a,b)\quad\iff\quad
\forall\,t\in(a,b)\,,\quad \dot\psi(t)\in H^1(\RE^3)\,,
$$
where $\psi$ is the unique solution of 
$\ddot{\psi}(t)  =   \Delta \psi(t) +\xi(t)\, G_i^1$ with zero
intitial data.
\end{lemma}
\begin{proof} Since the unique solution of $\ddot{\varphi}(t)  
=   \Delta \varphi(t) +\xi(t)\, \delta_{y_i}$ with zero
intitial data (at time $t=0$) is given by
$$
\varphi\,(t,x)=\frac{\theta(|t|-|x-y_i|)}{4\pi|x-y_i|}\, 
\xi(t-\sgn(t)\,|x-y_i|)$$ 
and $G_i^1=(-\Delta+1)^{-1}\delta_{y_i}$, we have
$(-\Delta+1)\psi=\varphi$. Thus 
\begin{align*}
\|(-\Delta+1)\psi(t)\|^2_{L^2}
=&\frac{1}{4\pi}\int_0^{|t|}dr\,|\xi(t-\sgn(t)r)|^2\\
=&\frac{1}{4\pi}\begin{cases}
\int_0^t ds\,|\xi(s)|^2\,,&t>0\\
\int_t^0 ds\,|\xi(s)|^2\,,&t<0
\,.\end{cases}
\end{align*}
Since, by Fourier transform (we suppose $t>0$, the case $t<0$ is analogous),
$$
\sqrt{|k|^2+1}\,\dot{\hat\psi}(t)=\frac{1}{(2\pi)^{3/2}}\,
\frac{1}{\sqrt{|k|^2+1}}\int_0^t ds\,\xi(s)\cos(t-s)|k|,  
$$
one has
\begin{align*}
&\|\sqrt{-\Delta+1}\,\dot\psi(t)\|^2_{L^2}\\
=&\frac{1}{2\pi^2}\,\lim_{R\uparrow\infty}
\int_0^t\int_0^tds\,ds'\,\bar\xi(s)\xi(s')\int_0^R
dr\,\frac{r^2\cos(t-s)r\,\cos(t-s')r}{r^2+1}\\
&=\frac{1}{4\pi}\int_0^t ds\,|\xi(s)|^2+\frac{1}{8\pi}
\int_0^t\int_0^tds\,ds'\,\bar\xi(s)\xi(s')
\left(e^{-|s-s'|}+e^{-2t}e^{-(s+s')}\right)\\
&\le \left(\frac{1}{4\pi}+t^2\,\frac{1+e^{-2t}}{8\pi}\right)\,
\int_0^t ds\,|\xi(s)|^2\,.
\end{align*}
Conversely $\|\sqrt{-\Delta+1}\,\dot\psi(t)\|^2_{L^2}$ for all
$t\in(a,b)$ implies 
$\xi\in L^2_{loc}(a,b)$ since the second term in the last equality
above is positive.
\end{proof}
\begin{lemma} Let $\varphi_i$ the solutions of the free wave equation
with initial data $\varphi_i(0)=\zeta_i G_i$,
$\dot\varphi_i(0)=\dot\zeta_i G_i$. Then
$$
\varphi_i(t,y_i)=\frac{\sgn(t)}{4\pi}\,\dot\zeta_i\,.
$$
\end{lemma}
\begin{proof}
Since $\psi(t):=\varphi_i(t)-(\zeta_i+t\dot\zeta_i)\,G_i$ satisfies 
$$\ddot\psi(t)=\Delta\psi(t)-(\zeta_i+t\dot\zeta_i)\,\delta_{y_i}$$
with zero initial data, one obtains
\begin{align*}
\varphi_i(t,x)=&
-\frac{\theta(|t|-|x-y_i|)(\zeta_i+(t-\sgn(t)|x-y_i|)\dot\zeta_i)}
{4\pi|x-y_i|}\\
&+\frac{\zeta_i+t\dot\zeta_i}{4\pi|x-y_i|}\,,
\end{align*}
and the proof is done by taking the limit $x\to y_i$.
\end{proof}
\begin{lemma} Let $\varphi$ the solution of the free wave equation with
regular initial data $\varphi(0)\in \bar H^2(\RE^3)$ and
$\dot\varphi(0)\in \bar H^1(\RE^3)$. Then for all $y\in \RE^3$ 
there exists $\zeta_y\in
C^1(\RE)$ with $\ddot\zeta_y\in L^2_{loc}(\RE)$ such that 
$$
\varphi(t,y)=\frac{\sgn
(t)}{4\pi}\,(\dot\zeta_y(t)-\dot\zeta_y(0))+\zeta_y(t)\,.
$$
Moreover $$\lim_{|t|\uparrow\infty}\,\varphi(t,y)=0\,.$$
\end{lemma}
\begin{proof} Let us consider the linear operator $\Delta_{1,y}$
corresponding to $Y=\left\{y\right\}$ and $V=1$. Then, 
by the results in \cite{[NP1]}, section 3 (also see \cite{[KP]},
theorem 3), Theorem 3.1 holds true for $\Delta_{1,y}$, 
with $\phi\in
C^0(\RE,D_{1,y})\cap C^1(\RE,\dot D_y)\cap C^2(\RE,L^2(\RE))$. 
This implies that, by Lemma 2.3, $\psi(t):=\phi(t)-\zeta^\phi(t)\, G_y^1$
belongs to $H^2(\RE^3)$ for all $t$ and that $\zeta_y\equiv\zeta^\phi\in
C^1(\RE)$ since $\dot D_y$ is normed by
$\|\phi\|^2_{\dot D_y}:=\|\nabla\phi_{reg}\|_{L^2}^2
+|\zeta^\phi|^2$. Since $\psi(t)$ solves the equation 
$\ddot\psi(t)=\Delta\psi(t)-\ddot\zeta_y(t)\,G_y^1$
with initial data $\psi(0)\in H^2(\RE^3)$ and $\dot\psi(0)\in
H^1(\RE^3)$, $\ddot\zeta_y\in L^2_{loc}(\RE)$ by Lemma 3.2. 
Moreover $\zeta_y$ solves the differential equation
$$\frac{\sgn(t)}{4\pi}\,\dot\zeta_y(t)+\zeta_y(t)=\phi_f(t,y)
$$  
so that, by Lemma 3.3,
$$
\varphi(t,y)=\frac{\sgn
(t)}{4\pi}\,(\dot\zeta_y(t)-\dot\zeta_y(0))+\zeta_y(t)\,.
$$
The fact that $\varphi(t,y)\to 0$ as $|t|\uparrow\infty$ follows from
the well known decay properties of the solution of the free wave
equation with regualar initial data.
\end{proof}
\begin{remark} The two previous lemmata show that $\phi_f(t,y_j)$ in 
(\ref{bc}) is made
of two pieces: a continuous and bounded one and another which has a
jump of size 
$\frac{\zeta^{\dot\phi_0}}{2\pi}$ at the origin. 
Thus, by taking the limit $t\to\pm 0$ in (\ref{delay}), $\dot\zeta(0_-)=\dot\zeta(0_+)=\zeta^{\dot\phi_0}$ and the
forward and backward solutions match together at the
initial time. 
\end{remark}
\vskip 10pt\p
{\it Proof of Theorem 3.1.} Let 
$$\psi(t):=\phi(t)-\sum_{1\le i\le n}\zeta_i(t-s)\, G_i^1\,.$$ Since $\zeta$
solves (\ref{delay}) and $\phi_f(\cdot,y_i)$ is a.e. derivable 
with a derivative in $L^2_{loc}(\RE)$ by Lemma 3.3 and Lemma 3.4, one has 
that $\zeta$ is piecewise $C^1$ with $\dot\zeta\in L^\infty(-T,T)
$ and $\ddot\zeta\in
L^2_{loc}((-T,T))$. 
Thus
$\psi(t)\in H^2(\RE^3)$ and $\dot\psi(t)\in H^1(\RE^3)$ for all 
$t\in (-T+s,T+s)$ by Lemma 3.2 since 
$$\ddot\psi(t)=\Delta\psi(t)-\sum_{1\le i\le n}\ddot\zeta_i(t-s)\,
G_i^1\,.$$ Since $G_i^1-G_i\in\bar H^2(\RE^3)$, this implies 
$\phi_{reg}(t)\in \bar H^2(\RE^3)$ and 
$\dot\phi_{reg}\in\bar H^1(\RE^3)$, where 
$$\phi_{reg}(t):=\phi(t)-\sum_{1\le i\le n}\zeta_i(t-s)\,
G_i$$ and $$\dot\phi_{reg}(t):=\dot\phi(t)-\sum_{1\le i\le n}\dot\zeta_i(t-s)\,
G_i\,$$ Thus $\dot\phi(t)\in\dot D_Y$ and
$\zeta^{\dot\phi(t)}=\dot\zeta(t-s)$. Moreover $\phi(t)\in
D_{V,Y}$, with $\zeta^\phi(t)\equiv\zeta(t-s)$ if the boundary
conditions (\ref{bc}) hold true for all $t\in (-T+s,T+s)$.
Since $\zeta$
solves (\ref{delay}) one has
\begin{align*}
&\lim_{x \to y_j}\, \left( \phi(t,x)-\zeta_{j}(t-s) G_{j}\right)\\=&
\phi_f(t-s,y_j)+ \sum_{i\neq j}\frac{\theta(|t-s|-|y_i-y_j|)}{4\pi|y_j-y_i|} \,
\zeta_i((t-s)-\sgn(t-s)|y_i-y_j|)\\
&+\frac{\zeta_i((t-s)-\sgn(t-s)|x-y_j|)-\zeta_j(t-s)}{4\pi|x-y_j|}\\=
&\phi_f(t-s,y_j) + \sum_{i\neq j}\frac{\theta(|t-s|-|y_i-y_j|)}{4\pi|y_j-y_i|} \,
\zeta_i((t-s)-\sgn(t-s)|y_i-y_j|)\\&-\frac{\sgn(t-s)}{4\pi}\,\dot\zeta_j(t-s)
=V_j(\zeta(t-s))
\end{align*}
and (\ref{bc}) are satisfied. Once we know that $\phi(t)\in D_{V,Y}$,
one has
$$
\ddot\phi=\Delta\phi+\sum_{1\le j\le n}\zeta_j\,\delta_{y_j}=
\Delta(\phi-\sum_{1\le j\le n}\zeta_j\,G_j)\equiv\Delta_{V,Y}\phi_{reg}
$$
and so $\phi$ solves (\ref{CP}).\par
Suppose now that $\varphi$ is another strong solution of (\ref{CP}). 
Then, by reversing the above argument, the boundary
conditions (\ref{bc}) imply that $\zeta^\phi$ solves the Cauchy problem
(\ref{delay}). By unicity of the solution of (\ref{delay}) one obtains
$\zeta^\phi(t)=\zeta(t-s)$. Then, defining
$$
\varphi_f(t):=\varphi(t)-\sum_{1\le j\le n}\phi_j(t-s)\,,
$$ 
where
$$
\phi_j(t,x):=\frac{\theta(|t|-|x-y_j|)}{4\pi|x-y_j|}\, 
\zeta_j(t-\sgn(t)\,|x-y_j|)\,,
$$
one obtains
$$
\ddot\varphi_f=\Delta\varphi_{reg}-\sum_{1\le j\le n}
(\Delta\phi_j+\zeta_j\delta_{y_j})=
\Delta(\varphi_{reg}-\sum_{1\le j\le n}
(\phi_j-\zeta_jG_j))=\Delta\varphi_f\,,
$$
i.e $\varphi_f$ solves the Cauchy problem (\ref{free}). 
Thus, by unicity of the solution of 
(\ref{free}), $\varphi=\phi$.\par 
The proof of (\ref{flow}) is standard: by considering the first components 
of $U_{V,Y}(t)U_{V,Y}(t_1)
(\phi_0,\dot\phi_0)$ and $U_{V,Y}(t+t_1)(\phi_0,\dot\phi_0)$ (with
$t\in [0,t_2]$)
one obtains two strong solutions of (\ref{CP}). They coincide by
unicity and so (\ref{flow}) holds true.
\qed
\begin{remark} By proceeding as in the linear case (see \cite{[KP]})
one can show that the wave equation $\ddot\phi=\Delta_{V,Y}\phi$ has
finite speed of propagation if and only if $V_j(\zeta)=V_j(\zeta_j)$ 
for all $j$.
\end{remark}
In the case the vector field $V$ is of gradient type, the flow
$U_{V,Y}(t)$ preserves an energy-like quantity:
\begin{lemma}
If $V=\nabla h$ then 
$$
\forall\,t\in (-T,T)\,,\qquad {\mathcal E}_{V,Y}U_{V,Y}(t)={\mathcal E}_{V,Y}\,,
$$
where the energy ${\mathcal E}_{V,Y}$ is defined by
$$
{\mathcal E}_{V,Y}(\phi,\dot\phi):=
\frac{1}{2}\,\left(\,\|\dot\phi\|^2_{L^2}+\|\nabla\phi_{reg}\|^2_{L^2}-(M_Y\zeta^\phi,\zeta^\phi)\,\right)
+{\text {\rm Re}}(h(\zeta^\phi))\,.
$$
\end{lemma}
\begin{proof} 
\begin{align*}
&\frac{d}{dt}\,\|\dot\phi\|^2_2=\langle\Delta_{Y,V}\phi,\dot\phi\rangle
+\langle\dot\phi,\Delta_{Y,V}\phi\rangle\\
=&\langle\Delta\phi_{reg},\dot\phi_{reg}+\sum_{1\le i\le
n}\dot\zeta^\phi_iG_i\rangle
+\langle\dot\phi_{reg}
+\sum_{1\le i\le n}{\dot{\zeta}^\phi_i}G_i,\Delta\phi_{reg}\rangle\\
=&\langle\Delta\phi_{reg},\dot\phi_{reg}\rangle-\sum_{1\le i\le
n}\dot\zeta^\phi_i\bar\phi_{reg}(y_i)
+\langle\dot\phi_{reg},\Delta\phi_{reg}
-\sum_{1\le i\le n}{\dot{\bar\zeta}^\phi_i}\phi_{reg}(y_i)\\
=&\langle\Delta\phi_{reg},\dot\phi_{reg}\rangle
+(M_Y\zeta^\phi,\dot\zeta^\phi)
-\sum_{1\le i\le
n}\dot\zeta^\phi_i\bar V_i(\zeta^\phi)\\
&+\langle\dot\phi_{reg},\Delta\phi_{reg}\rangle+(\dot\zeta^\phi,M_Y\zeta^\phi)
-\sum_{1\le i\le n}{\dot{\bar\zeta}^\phi_i} V_i(\zeta^\phi)\\
=&\frac{d}{dt}\,
\left(\,-\|\nabla\phi_{reg}\|^2_2+(M_Y\zeta^\phi,\zeta^\phi)
-2{\text {\rm Re}}(h(\zeta^\phi))\,\right)\,.
\end{align*}
\end{proof}
The above conservation result can be used to obtain a global existence
result by standard arguments:
\begin{theorem} Let $V=\Theta+\nabla h$ with $\Theta$ an Hermitean
matrix and $h$ such that 
$$
{\text{\rm Re}}(h(\zeta))\ge c_1\,|\zeta|^2-c_2\,,
\quad c_1\ge 0\,,\ c_2\ge 0\,.
$$
Then the flow $U_{V,Y}(t)$ is global.
\end{theorem}
\begin{proof} In this case ${\mathcal E}_{V,Y}(\phi,\dot\phi)
=\frac{1}{2}\,(\|\dot\phi\|^2_2+
{\mathcal F}_{\Theta,Y}(\phi))+{\text{\rm Re}}(h(\zeta))$. 
Since $V$ is of gradient type and ${\mathcal F}_{\Theta,Y}$ is bounded
from 
below (see
\cite{[AGHH]}), 
$$
|\zeta^\phi(t)|^2\le k\,{\mathcal E}_{V,Y}(\phi(t),\dot\phi(t))=
k\,{\mathcal E}_{V,Y}(\phi(0),\dot\phi(0))
$$
for some positive constant $k$
\end{proof}

\section{blowing up solutions and their lifespan}

The solution given in Theorem 3.1 can be oviously extended to
the space-time domain
$$
E=\bigcap_{y\in Y}\left\{(t,x)\in \RE^4\,:\,
-T+s-|x-y|<t<T+s+|x-y|\,\right\}\,.
$$
Such a function is a local solution on $E$ in the
sense that
\begin{align*}
&\partial^2_{tt}\phi(t,x)= 
\Delta\left(\phi-\sum_{1\le j\le n}\zeta_j^\phi G_j\right)(t,x)
\qquad(t,x)\in E\\
&\phi(s) =  \phi_0 \\
&\dot\phi(s)  =  \dot\phi_0\,. 
\end{align*}
Note however that no boundary conditions can be imposed on $\phi$ at times
$t\notin (-T+s,T+s)$, since $\left\{t\right\}\times Y$ is not included
in $E$ when $t\notin (-T+s,T+s)$. \par
If $\zeta(t)$ blows up at times $\pm T$ then the above local
solution has a blow-up boundary given by $\partial E$.
\par
Now we turn to the detailed analysis of the case in which 
$Y=\left\{y\right\}$, so that
there is no delay in (\ref{delay}). Since (by direct verification) 
the backward solution $\zeta_-$ of 
(\ref{delay}) is related to a
forward solution by $\zeta_-(t)=\zeta^-_+(-t)$, where $\zeta^-_+$ is
the forward solution of (\ref{delay}) with inhomogeneous term
$\phi_f(-t,y)$, we
will concentrate the analysis on the solutions of the system
\begin{align*}
&\dot \zeta(t)+V(\zeta(t))= g(t)\\
&\zeta(0)=\zeta_0   
\end{align*}
where $g(t)$ is continuous and $g(t)=g_0+g_1(t)$ with $g_1(t)\to 0$ as
 $|t|\uparrow \infty$ (see Remark 2.5). We will study the
 equation in a real framework, i.e. we suppose that the fields $\phi$
 and $V$ are
 real-valued. Hence $\zeta(t)\in\RE$. Moreover, to fix ideas, let us consider 
a continuous function $V:\RE \to \RE$, and regular enough to assure local 
existence and unicity of the solution of the differential equation.
\par
Let us begin by considering preliminarly the case in which $g(t)=g_0$ is constant. 
This gives the autonomous differential equation $\dot z+V(z)=g_0$ with 
equilibrium (constant) solutions given by the $z$'s which satisfy the 
equation $V(z)=g_0$.
Let us fix an initial datum $z_0=z(0)$ not belonging to such a set. 
Correspondingly, in the interval of existence of the solution 
the term $g_0-V(z(t))$ has constant sign by continuity, and the solution  
$z(t)$ is implicitly given by the relation 
$$
\int_{z_0}^{z(t)}\frac{ds}{g_0-V(s)} = t\,.
$$
This implies an elementary but fundamental remark. The solution of the 
auxiliary equation is global if and only if both the improper integrals 
$$
\int_{z_0}^{\pm \infty}\frac{ds}{g_0-V(s)} 
$$
diverge. If, on the contrary, at least one of them converges, the
solution blows up in the past or in the future and the backward or
forward lifespan $T_\pm$  of the solution 
is given just by the value of one of such integrals.  
\par
Now the main point is to include in the analysis the time dependent
term $g_1(t)$. The presence of this term is an essential preclusion to
the possibility of writing down a closed formula for the solution of
our differential equation, and one has to make resort to other
methods.  A first remark is that the time dependent term $g_1(t)$ 
is bounded.  This suggests that the behaviour of the solution of the
inhomogeneous equation could be not affected so much by this term.
The idea is to use differential inequalities to confront the size of
solutions of the two equations. 
Roughly speaking, a solution greater or lower than a function blowing up to 
$+ \infty$ or $-\infty$ respectively is blowing up;
and a solution which is bounded between two funtions finite at every finite 
time is global in time. Both situation occur, and both can occur for the same coupling 
depending on the initial data.
The analysis is based on differential inequalities which relate the 
solution of a comparison 
auxiliary equation with the solution of the given equation.
We recall briefly that the defect operator $P$ associated to the
differential equation $\dot z(t)=F(t,z(t))$
is given by $$P(t,z):=\dot z -F(t,z(t))\,.$$ 
Comparison of defect operator leads to important and classical
differential 
inequalities (see e.g. \cite{[Walter]}):
\begin{theorem}
 Let $z_-(0)\leq z(0)\leq z_+(0)$ and 
$$P(t,z_-)\leq 0=P(t,z)\leq P(t,z_+)\,,\quad t\in[a,b]\,.$$ 
Then one has $z_-(t)\leq z(t)\leq z_+(t)$ in $[a,b]$.  
\end{theorem}
Correspondingly, with a terminology introduced by Perron, $z_-$ is
called subsolution and $z_+$ is called supersolution.\par  
Now, let us define $K:=\sup_{t\in\RE}|g(t)|$ and consider the couple of 
differential equations
$$
\dot z_\pm(t)+V(z_\pm(t))=\pm K
$$
with initial conditions $z_-(0)\le \zeta_0$ and $z_+(0)\ge \zeta_0$.
It is immediate to see that one has the following 
inequalities between defect operator:
$$
P(z_-,t)=g(t)-K\leq 0=P(t,\zeta)\leq g(t)+K=P(t,z_+)\,.
$$ 
So $z_-$ is a subsolution and $z_+$ is a supersolution of $\zeta$. 
Of course to a subsolution $z_-$ positively blowing up in the future 
with a lifespan
$T_+$ corresponds a solution $\zeta$ positively blowing up with a 
lifespan $T_*<T_+$. A similar reasoning applies to negatively blowing
up supersolutions. Since by Theorem 3.1 $U_{V,Y}(-t)=U_{V,Y}(t)^{-1}$,
we do not take care of solutions blowing up in the past. Summarizing 
we obtain the following
criterium for global existence or blowup:
\begin{theorem}
Let $\phi(t)$ be the solution of the Cauchy problem (\ref{CP}) with 
$\zeta_0=\zeta^{\phi(0)}$ and put $K:=\sup_{t\in\RE}|\phi_f(t,y)|$. 
{\item 1)} 
$\phi$ is a global solution if the integrals 
$$
\int_{\zeta_0}^{\pm\infty}\frac{ds}{K +V(s)}\,,\quad
\int_{\zeta_0}^{\pm\infty}\frac{ds}{K-V(s)}   
$$
diverge;
{\item 2)}
$\phi$ is positively blowing up in the
future if 
$$
-\int_{\zeta_0}^{+\infty}\frac{ds}{K+V(s)} 
$$
converges to a positive value. The value of such an integral 
gives an upper bound of
the lifespan $T_*$. An analogous statement holds true for solutions negatively blowing up 
in the future.
\end{theorem}
\begin{remark}
Note that the constant $K$ depends on both $\phi(0)$ and $\dot
\phi(0)$. So as it should be expected, the complete set of initial
data determines the global existence or blow up
of the solution.
\end{remark}
\begin{remark} Another very simple criterium of global existence is the
following: suppose that both the sets $S_\pm=\{s_\pm: V(s_\pm)=\pm K\}$
are not void. Then
any $s_\pm\in S_\pm$ provide global (stationary) super and subsolutions. 
This gives global existence for solution with $s_-\le \zeta_0\le s_+$.
\end{remark}
\subsection{Examples} Tipical nonlinearities in model equations
are given by power law, or more generally polynomial couplings. 
They are essentially to be
considered as phenomenological choices, tipically originated by some
ad hoc truncation of a Taylor approximation of more general
couplings. \par
Let us consider the function
$$
V(\zeta)=\gamma|\zeta|^{\sigma}\zeta, \quad \quad \gamma>0\,,\quad 
\sigma \in \RE\,.
$$
The auxiliary equations 
$$
\dot z_\pm=-\gamma|z_\pm|^{\sigma}z_\pm\pm K\,
$$
have the equilibrium solutions 
$s_\pm=\pm \left(\frac{K}{\gamma}\right)^{\frac{1}{\sigma+1}}$. 
Thus, by Remark 4.4, one has a global solution for any initial data 
with $|\zeta_0|\le \left(\frac{K}{\gamma}\right)^{\frac{1}{\sigma+1}}$.
For data with $|\zeta_0|>
\left(\frac{K}{\gamma}\right)^{\frac{1}{\sigma+1}}$ and for $\sigma>0$,
the integral
$$
\int_{\zeta_0}^{\infty}\frac{ds}{\gamma|s|^{\sigma}s+K}
$$
converges and so in this case we have blow up with a lifespan 
$$
|T_*|<\int_{\zeta_0}^{\infty}\frac{ds}{\gamma|s|^{\sigma}s+K}\,.
$$ 
A nonlinearity which deserves attention is given by $V(\zeta)=\alpha \zeta^2$. This admits an
analysis analogous to the one just devised and corresponds to a
quadratic nonlinearity in the abstract wave equation. The peculiarity
is that in this case (\ref{delay} ) is a Riccati equation, one
of the better known nonlinear differential equations of the first
order and one with wide applications in mathematics and sciences. One
of the more striking properties of the Riccati equation is that by a
nonlinear transformation of the unknown function, it can be reduced to
a second order linear differential equation, and this fact appears as
particularly noteworthy in our context, where the original problem is
a wave equation with a quadratic (concentrated)
nonlinearity. We are not able, till now, to judge about the 
relevance of this fact, which seems to deserve further investigation.
Another well known property of Riccati equation is
the fact it has always at least one nonglobal solution (see e.g. 
\cite{[Hille]}) when the time dependent term $g(t)$ is an
algebraic function. Of course, there is no hope that the evaluation at
$y$ of a solution of a free wave equation be an algebraic equation, 
at least for generic data, but
in our case, thanks to the properties of $\phi_f(t,y)$ the
simple majorizations above allow to obtain blowing up solution also in
the case of nonalgebraic inhomogeneous terms.  Moreover, another
important fact about blowing up solutions of the Riccati equation is
the typical behaviour of the solution in the proximity of the blow up
time $T_*$ lifespan, which is of the type
$$\zeta(t)\sim \frac{1}{t-T_*}\,.
$$ This gives, in view of the relation between the time behaviour of
$\zeta(t)$ and the behaviour of the solution, the qualitative
asymptotic spacetime behaviour of the solution of the wave equation
with quadratic concentrated nonlinearity near the blow up time, 
which is of the type
$$
\phi(t,x)\sim \frac{1}{t-T_*-|x-y|}\,\frac{1}{|x-y|}\,.
$$
Similar consideration hold for the power nonlinearities analyzed above, or
other non polynomial couplings for which a precise analysis of the
equation for $\zeta(t)$ is feasible.

\end{document}